# Detecting Magnetic Ink Barcodes with Handheld Magnetoresistive Sensors

Sofia Abrunhosa[1,2], Ian Gibb[3], Rita Macedo[1], Emrys Williams[3], Nathalie Muller[3], Paulo P. Freitas[1,2], Susana Cardoso[1,2]

[1]Instituto de Engenharia de Sistemas e Computadores – Microsistemas e Nanotecnologias, Lisbon, Portugal
[2]Universidade de Lisboa, Instituto Superior Técnico, Lisbon, Portugal
[3]MagVision, Frome, United Kingdom

**Information encoding in barcodes using magnetic-based technology is a unique strategy to read data buried underneath non-transparent surfaces since a direct line-of-sight between the code and the reader is not required. This technology is of particular interest in secure labelling and recyclable packaging applications. However, current magnetic reading heads, such as those employed for magnetic ink character recognition, need to be placed in contact with the magnetic structures, limiting the depths at which the information can be read. This paper describes a strategy to overcome that limitation by replacing the traditional inductive heads with tunnel magnetoresistive (TMR) sensors. Soft-magnetic codes can be printed using conventional LaserJet toners and, by having their magnetisation set with a permanent magnet included in the device, the resulting magnetic field can be read using a TMR sensor. We demonstrate that such a device can read barcodes at depths of at least 1 mm. It can also resolve individual structures as thin as 200 μm when used in contact.**

## I. Introduction

LABELLING objects with optical barcodes is an established technology. The codes are comprised of ink structures with varying width and separation, allowing the encoding of digital information related to the item they are attached to. These codes can then be read using commercially available optical readers to retrieve the information [1]. However, any damage to the surface or obstruction of the reading path limits the readability of a given barcode. As a result, applications where the information must be contained underneath the surface could make use of a no line-of-sight technology such as magnetic-based barcodes, which are more accessible and scalable than, for example, radio-frequency identification tags [2][3]. This paper discusses the possibility of using ferromagnetic printable ink or toner, which is readily available and can be read using magnetic-based sensors.

The usage of magnetic ink character recognition (MICR) technology is already widespread in the banking industry [4]. Conventional cheques usually include information written in alphanumeric characters printed with hard-magnetic ink and read with inductive heads [5]. However, resorting to inductive heads limits the applications to in-contact reading, which makes it unfeasible to read information buried underneath at depths higher than a few hundreds of μm.

Given the small magnetic fields created by these magnetic barcodes (<3 Oe) and the need to use structures with small widths (1-0.1 mm) for increased bit density, a sensor technology with high sensitivity and good spatial resolution is required. Tunnel magnetoresistive (TMR) sensors appear as an ideal candidate, due to their low detectivities (down to pTesla) and small footprints (few μm) [6][7][8]. Implementations of these sensors for magnetic encoders using soft magnetic media have already been demonstrated [9][10].

In this paper we describe a portable device containing TMR sensors that can detect and decode the magnetic field created by soft-ferromagnetic barcodes printed with conventional LaserJet toners. To increase the magnetisation of the medium while it is being scanned, the system also contains a disc-shaped permanent magnet placed behind the sensor. However, the presence of this magnet will also induce stray fields in the vicinity of the TMR sensor, which can affect its sensitivity. Therefore, a fine balance must be found between maximising the magnetisation of the barcode while also limiting the effect on the sensor transfer curve.

## II. Simulation Model for the Magnetic Fields

To obtain the magnetic field created by the barcode, each stripe is modelled with a uniform out-of-plane (z axis) magnetisation across the whole structure and a uniform geometry along its length. Therefore, a 2D approach can be considered [11]. For a single stripe, the in-plane (x-axis) of the resulting magnetic field can be described as:

$$H_x^{stripe}(x,z) = \frac{M}{2\pi}\left(\log\frac{r_4}{r_2} - \log\frac{r_3}{r_1}\right) \quad (1)$$

$M$ is the magnetization and $r_k$ are the distances between the measurement point and each of the four geometric corners of the structure's cross-section. The overall magnetic field is then obtained by linear superposition of the field created by each structure.

The magnetic field created by a permanent disc magnet can be computed by evaluating a surface integral at both the top and the bottom circles for an axially magnetised magnet, using equations (2) to (4) at heights $z$ and $z+t$ and assuming that $M(z) = -M(z+t)$, for a given magnet thickness $t$.

$$H_x(x,y,z) = \frac{M}{4\pi}\int_0^D\int_0^{2\pi} d\rho d\theta \frac{\rho(x-\rho\cos\theta)}{\sqrt{(x-\rho\cos\theta)^2+(y-\rho\sin\theta)^2+z^2}} \quad (2)$$

$$H_y(x,y,z) = \frac{M}{4\pi}\int_0^D\int_0^{2\pi} d\rho d\theta \frac{\rho(y-\rho\cos\theta)}{\sqrt{(x-\rho\cos\theta)^2+(y-\rho\sin\theta)^2+z^2}} \quad (3)$$

$$H_z(x,y,z) = \frac{M}{4\pi}\int_0^D\int_0^{2\pi} d\rho d\theta \frac{\rho z}{\sqrt{(x-\rho\cos\theta)^2+(y-\rho\sin\theta)^2+z^2}} \quad (4)$$

$x, y, z$ are the scalar components of the position vector where the magnetic field is calculated and $D$ is the diameter of a disc magnet.

### III. MAGNETORESISTIVE SENSOR READING HEAD

The proposed device (Fig. 1) is comprised of two major components: a chip containing two TMR sensors placed at the edge of the system to minimise the distance to the barcode (≈200 μm when reading in contact) along with an axially magnetised disc-shaped permanent magnet placed behind the chip and used to maximise the magnetisation of the soft-ferromagnetic structures during scanning. The sensors are sensitive to the in-plane (x axis) component of the field while the permanent magnet creates a magnetic field predominantly out-of-plane (z axis) at barcode level. Due to the small dimensions of the device, since its final objective is to be used in handheld applications, electrical connections from the sensors to the external acquisition electronics are achieved using a flexible printed circuit (FPC). Finally, a 3D printed housing holds all the components and is shaped so that the distance between the sensors and the barcode is minimised during handheld scans.

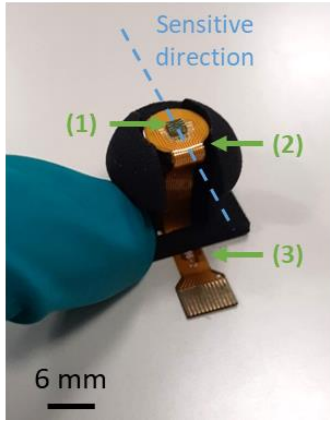

Fig. 1. Photo of magnetic reading head after assembly. (1) chip containing TMR sensors; (2) disc magnet (not visible); (3) flexible printed circuit. The sensors are sensitive along the direction represented by the dashed blue line. Before operation, the chip and wiring are coated with a 200 μm thick epoxy layer for mechanical protection.

#### A. TMR sensor

The TMR stack was deposited using a Nordiko 3600/8800 magnetron sputtering system [12] with the following composition: [Ta 5 / CuN 15]x3 / Ru 5 / MnIr 8 / 2.2 CoFe / Ru 0.7 / CoFeB 1.9 / MgO 1.5 / CoFeB 3 / Ru 0.2 / NiFe 4 / Ta 10 / Ru 10 (thicknesses in nm). The magnetic tunnel junctions were patterned by direct-write laser lithography and ion milling etching. The pillar passivation was done using physically enhanced chemical vapour deposition and vias to the top electrode were opened by reactive ion etching. The metal for external contacts was deposited through magnetron sputtering and etched using reactive ion etching as well.

Each TMR sensor was comprised of 20 magnetic tunnel junctions in series, with individual dimensions equal to 2 μm × 20 μm. A single sensor occupied a total sensitive area of 70 μm × 128 μm. The process resulted in TMR sensors with $R_{min}$≈250 Ω (resistance-area product $R_{min} \times A \approx 0.5$ Ω·μm$^2$) and 80-90% TMR. Each chip contained two TMR sensors, occupied a volume of 2 mm × 2 mm × 0.8 mm and was placed horizontally with respect to the barcode to confer an in-plane sensitive direction to the device. Any magnetic fields placed along this direction can be detected by the sensor, while it remains insensitive to out-of-plane fields.

#### B. Soft-ferromagnetic barcode

The barcode structures were printed with an office LaserJet printer using HP 402 toner, containing iron oxide nanoparticles that confer a soft-ferromagnetic behaviour to the printed structures. This was confirmed by measuring its M(H) curve on a vibrating sample magnetometer, which showed a saturation field of 1250 Oe and a saturation magnetisation of 15 emu/cm$^3$. Therefore, to maximise the signal output of the reading head, it is necessary that the permanent magnet produces a field equal to or higher than this value at the desired reading distances.

Solving equation (1) for a given barcode gives the signal pattern represented in Fig. 2, which corresponds to the in-plane component of the magnetic field that the TMR sensors are sensitive to.

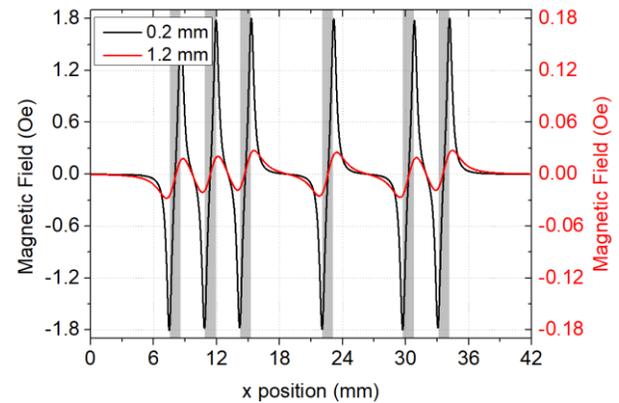

Fig. 2. Simulation of the x-component of the magnetic field created by a barcode magnetised along the z-axis, for two different reading distances: 0.2 and 1.2 mm. The barcode included structures with 1.1 mm width, separated by 2.3 mm or 6.8 mm, each with a thickness of 20 μm and a magnetisation of 15 emu/cm$^3$. For clarity purposes, the y-axis scale for the signal read at 1.2 mm was decreased by a factor of 10.

The maxima and minima of the signal correspond to the edges of the structures. Therefore, by measuring the peak positions, these can be correlated to the position of each stripe and the code is reconstructed. This is also the approach taken to decode real signals scanned with the constructed reading head.

In an actual system, however, the sensors would not be completely flat against the barcode surface. One immediate effect would be the increase in reading distance due to the inclination, which would reduce the amplitude of the signal and the spatial resolution of the system. Additionally, the sensors would no longer solely detect the x-component of the field but also the z-component. While for the x-component the peaks correspond to the edges of the structures, for the z-component they correspond to their centres. This would cause a shift in the position of the peaks when compared to a system with no inclination. Nevertheless, we have observed that, with increasing tilt, the signal-to-noise ratio of the output is lost due to higher reading distance before any distortions to the signal become relevant.

*C. Permanent magnet*

The sensor head integrates a disc magnet with 8 mm of diameter and 2 mm of thickness. This was chosen as the configuration that guarantees readings at 1 mm of distance from the barcode while reducing the overall dimensions of the device that keep it portable.

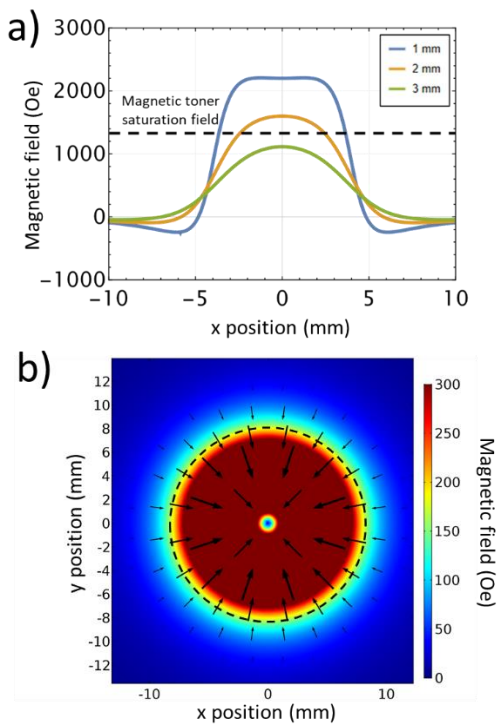

Fig. 3. Magnetic field created by a disc-shaped permanent magnet (diameter = 8 mm, thickness = 2 mm). (a) vertical component of the field at three different distances from the magnet (barcode level) together with the saturation field of the toner (1250 Oe). (b) radial component of the field created 0.8 mm from the magnet (sensor level), with the dashed black line representing the edge of the material. The TMR sensors are placed at the geometric centre, where the field strength is minimum.

The magnetic field created by the magnet was simulated using equations (2) to (4). Fig. 3a shows the out-of-plane component of the magnetic field created by the permanent magnet obtained along its diameter. This component, responsible for magnetising the barcode during scanning, has a maximum strength of 2.2 kOe at 1 mm depth, which is enough to guarantee the saturation of the structures without affecting the TMR sensor. This is true for distances up to 2.7 mm and beyond that value the system is not operating under ideal conditions. Fig. 3b shows the radial component of the field ($|H_x + H_y|$) at the sensor level (0.8 mm away from the magnet). This component of the field confers an offset to the transfer curve of the sensor and hinders its sensitivity in an effect akin to biasing TMR sensors with permanent magnets for increased linearity [13].

The region with the smallest field strength is close to the geometric centre of the magnet. If the centre of the TMR chip is perfectly aligned with the centre of the magnet, the field at the edges of the sensors (±128 μm away) is ≈40 Oe. This value is very close to the sensor saturation field (obtained from the R(H) transfer curve), which reduces the sensitivity by 20-30% of the initial sensitivity, with higher impact on the magnetic tunnel junctions closest to the periphery of the sensor. Fig. 4 shows a comparison between the transfer curve of the sensor after fabrication and that of the same sensor after assembly. The magnet-sensor alignment protocol in use ensures a placement precision of 50 μm by monitoring the sensitivity of the sensor during the assembly.

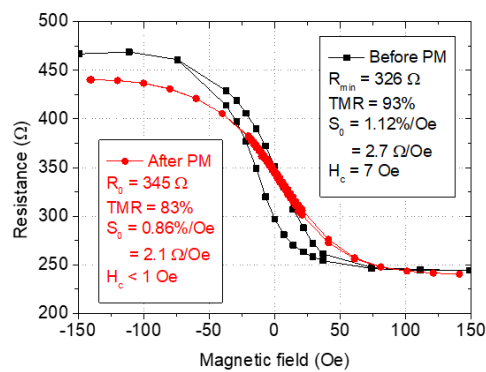

Fig. 4. TMR sensor R(H) curves before and after placement on the permanent magnet (PM). Despite the sensitivity decrease by 24%, which is detrimental to the application, the reduction of the coercive field and improved linearity in the centre region facilitate the processing of the resulting signals.

While this sensitivity decrease limits the capabilities of the system, improvements can be made by using larger permanent magnets (both in diameter and thickness). In those cases, the region where the sensor can be safely placed without severely reducing its sensitivity increases, and so can the strength of the field at barcode level for higher depths. Nonetheless, this approach limits the device's portability.

## IV. TECHNOLOGY VALIDATION

After assembly, the reading head is connected to a printed circuit board containing the electronics for sensor biasing and signal processing. The two sensors are connected in a half Wheatstone bridge with two external 330 Ω resistors and the output signal goes through a two-stage amplifier with a variable gain of 190-1600 times.

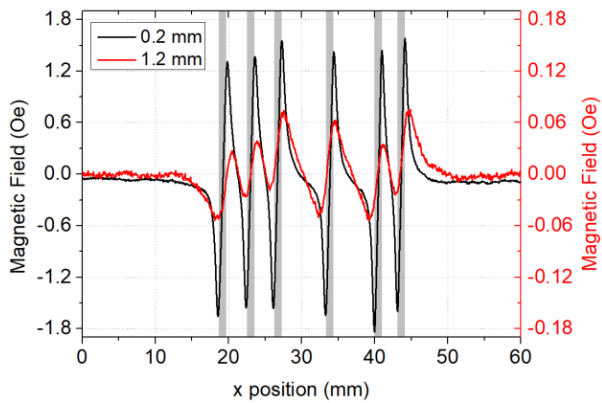

Fig. 5. Reading head output after scanning a ferromagnetic barcode at two different reading distances (distance from TMR sensor to the barcode). Two TMR sensors in a half Wheatstone bridge were used with a combined sensitivity of 19 Ω/Oe ($I_{bias}$ = 1 mA). Individual barcode stripes (1.1 mm wide) are represented as grey rectangles under the signal. For clarity purposes, the y-axis scale for the signal read at 1.2 mm was decreased by a factor of 10.

Fig. 5 contains the signal measured while scanning the same barcode with the reader in contact and 1 mm away. The amplifier gain setting was different for both cases: 410 for the signal read with the reader in contact and 1600 for the one read 1 mm away. This also results in a noisier output for the signal that is further away. The encoded information can be retrieved by finding the maxima and minima positions and correlating those with the expected structure positions. Other encodings with increased bit density are being explored. This reader can also resolve barcodes with structure widths as small as 200 µm, when used in contact.

## V. Conclusions

We have demonstrated a reading head based on tunnel magnetoresistive sensors capable of resolving 1.1 mm wide barcode structures at a reading distance of at least 1 mm. This is a promising step forward from the inductive head technology currently used to read information encoded in magnetic ink. Improvements of the system at all stages can still be achieved, the most significant of them being the increase in magnet size to minimise the effect on the sensitivity of the sensor. Moreover, the current approach can be scaled up to read multi-track barcodes or even QR codes by using several of the existing chips or having several pairs of sensors inside one longer chip.

## Acknowledgment

The authors acknowledge the Mag-ID project, funded by the European Commission Directorate General for Research & Development as part of the Horizon 2020 – research and innovation framework program under the Proposal ID H2020-FTI-870017. The authors also wish to acknowledge the Fundação para a Ciência e a Tecnologia for funding of the Research Unit INESC MN (UID/05367/2020) through plurianual BASE and PROGRAMATICO financing, the PhD research grant of S. Abrunhosa (PD/BD/150392/2019) and project LISBOA-01-0145-FEDER-031200.